\documentclass[showpacs,showkeys,unsortedaddress,preprint]{revtex4-1}

\usepackage[pdftex]{graphicx}
\usepackage{epstopdf}
\usepackage{amsbsy} 
\usepackage{xspace}
\usepackage{amsmath}
\usepackage{amssymb}

\usepackage{graphicx}%

\usepackage{accents}
\usepackage{color}
\usepackage{bbold}

\usepackage{makeidx}

\makeindex

\begin{document}

\title{Observability of surface Andreev bound states in a topological insulator in proximity to an s-wave superconductor}
\author{M. Snelder}
\affiliation{Faculty of Science and Technology and MESA+ Institute for Nanotechnology, University of Twente, 7500 AE Enschede, The Netherlands}
\author{A. A. Golubov}
\affiliation{Faculty of Science and Technology and MESA+ Institute for Nanotechnology, University of Twente, 7500 AE Enschede, The Netherlands}
\affiliation{Moscow Institute of Physics and Technology, Dolgoprudny, Moscow 141700, Russia}
\author{Y. Asano}
\affiliation{Department of Applied Physics and Center for Topological Science \& Technology, Hokkaido University, Sapporo 060-8628, Japan}
\author{A. Brinkman}
\affiliation{Faculty of Science and Technology and MESA+ Institute for Nanotechnology, University of Twente, 7500 AE Enschede, The Netherlands}

\date{\today}

\begin{abstract}
To guide experimental work on the search for Majorana zero-energy modes, we calculate the superconducting pairing symmetry of a three-dimensional topological insulator in combination with an $s$-wave superconductor. We show how the pairing symmetry changes across different topological regimes. We demonstrate that a dominant $p$-wave pairing relation is not sufficient to realize a Majorana zero-energy mode useful for quantum computation. Our main result is the relation between odd-frequency pairing and Majorana zero energy modes by using Green functions techniques in three-dimensional topological insulators in the so-called Majorana regime. We discuss thereafter how the pairing relations in the different regimes can be observed in the tunneling conductance of an $s$-wave proximized three-dimensional topological insulator. We discuss the necessity to incorporate a ferromagnetic insulator to localize the zero-energy bound state to the interface as a Majorana mode.  
\end{abstract}
\maketitle
The search for a qubit that is robust against local decoherence has led to extensive studies of the so-called Majorana zero-energy mode in materials with strong spin-orbit coupling (SOC) combined with $s$-wave superconductors. Nanowires and topological insulators are promising candidate materials with strong spin-orbit coupling \cite{Fu2008,Fu2009,Fu22009,Tanaka09,Akhmerov09,Linder10,Sau2010,Alicea2010,Oreg10,Lutchyn10,alicea11,qi11,Pikulin2012,Asano2013,Stanescu2013}. Experimentalists have already reported signatures of the Majorana zero-energy mode \cite{Mourik2012,Deng2012,Das2012}, where zero-bias conductance peaks are the main features observed in this context.

Besides the possibility to host  Majorana modes, topological materials are also interesting for the study of unconventional $p$-wave superconductivity by itself. $p$-wave superconductivity gained renewed interest after the prediction that the $p$-wave pairing symmetry in He$^{3}$ would lead to half-quantum vortices with  potential application to the field of quantum computing \cite{Osheroff1972,Osheroff1972b,Leggett}. Sr$_{2}$RuO$_{4}$ is believed to have $p$-wave symmetry and together with the prediction of the existence of a nodal gap it has led to an extensive study of this material \cite{Ishida1998,Luke1998,Afterberg1998,Kealey200,Duffy2000,Eremin2004,Kallin2009,Kallin2012}. Also superconductor/ferromagnet/superconductor junctions are studied for their prospective to switch from a dominant $s$-wave state to a $p$-wave state \cite{Ryazanov2001,Golubov2002,Matsuda2009,Samokhvalov2014,Buzdin2005,Bergeret2005}. Also these devices have potential application to quantum computation, through which the understanding of the behaviour of $p$-wave superconductivity has become a field on its own.  

In this paper we study the $s$ and $p$-wave correlations that exist in three-dimensional topological insulators (3D TIs). We first discuss in which regime a dominant $p$-wave correlation is present and under which conditions a dominant $p$-wave correlation can lead to a zero-energy Majorana bound state. We derived the expressions for the bound states formed at a 3D TI in the ``Majorana" regime. We show the relation between odd-frequency pairing and Majorana zero-energy modes by using a Green function approach. We use those new insights to calculate the conductance spectra of proximity induced superconducting 3D TIs with and without broken time-reversal symmetry. The main focus is to understand how the modelled tunneling conductance spectra arise from the $s$ and $p$-wave correlations. We will see that the combination of a zero-bias conductance peak together with conductance dips at the gap energy are distinguished features for $p$-wave correlations. We demonstrate that it is not necessary to be in the ``Majorana" regime in order to observe signatures of the $p$-wave correlations.

\section{Pairing wave function and Majorana-modes in a 3D TI}
In order to determine the pairing relations for a material with strong SOC we start with the relation $[E-H]\textbf{G}=\mathbb{1}$ where $\textbf{G}$ is the Green function of the system, $\mathbb{1}$ the identity matrix and $H$ is in the basis $\left[\psi_{\upharpoonleft},\psi_{\downharpoonleft},\psi^{\dagger}_{\upharpoonleft},\psi^{\dagger}_{\downharpoonleft}\right]$ \cite{Asano2013}. For a system with spin orbit coupling $\left[E-H\right]$ is given by
\begin{eqnarray}
\left[\begin{array}{cccc}
E+\epsilon_{p}+M & -\lambda_{x}k_{x}-i\lambda_{y} k_{y} & 0 &-\Delta \\
-\lambda_{x}k_{x}+i\lambda_{y} k_{y} & E+\epsilon_{p}-M & \Delta & 0\\
0 & \Delta & E-\epsilon_{p}-M & -\lambda_{x}k_{x}+i\lambda_{y} k_{y}\\
-\Delta & 0 & -\lambda_{x}k_{x}-i\lambda_{y} k_{y} & E-\epsilon_{p}+M
\end{array} \right],
\end{eqnarray} where $\epsilon_{p}=\mu-\dfrac{\hbar k^2}{2m}$. Here, $\lambda_{x,y}$ is the SOC strength in $x$ and $y$-direction. At the end we will set the SOC strength equal to the Fermi velocity and $\epsilon_{p}=\mu$ to discuss the superconducting correlations in the topological surface states. $\mu$ is the chemical potential, $\Delta$ is the induced superconducting gap and $M$ is the energy of a perpendicular magnetic field, either externally applied (Zeeman term) or due to an exchange interaction. Note, that the Zeeman contribution of a magnetic field can be considerable, due to the large $g$-factor of most relevant materials. We neglect the orbital contribution of the applied field. 

By taking the inverse of this matrix equation we can obtain the Green's function \textbf{G}, expressed as
\begin{eqnarray}
\left[\begin{array}{cc}
G_{11}& G_{12}\\
G_{21} & G_{22}
\end{array} \right].
\end{eqnarray} The elements $G_{ij}$ are blocks of 2$\times 2$ matrices. The diagonal blocks describe the propagation of the electrons and holes separately. The off-diagonal blocks describe the combined electron and hole propagation, i.e. they describe the propagation of the Cooper pairs in the $s$-wave proximized topological insulator which is the quantity of interest in this paper. When we represent
\begin{align}
E-H(\boldsymbol{k})=
\left[ \begin{array}{cc} \hat{a} & \hat{b} \\
 \hat{c} & \hat{d} \end{array} \right],
\end{align}
the Green function is given by
\begin{align}
\mbox{\textbf{G}}(\boldsymbol{k}, E)
=
\left[ \begin{array}{cc} ( \hat{a} - \hat{b}\; \hat{d}^{-1}\; \hat{c} )^{-1} 
& ( \hat{c} - \hat{d}\; \hat{b}^{-1}\; \hat{a} )^{-1} \\
( \hat{b} - \hat{a}\; \hat{c}^{-1}\; \hat{d} )^{-1} &
( \hat{d} - \hat{c}\; \hat{a}^{-1}\; \hat{b} )^{-1}
\end{array} \right].
\end{align}
We find 
\begin{align}
G_{12} 
=& i \frac{\Delta}{Z} \left[ 
(E^2 - B_{k,M}) \hat{\sigma}_0
+2 (- \lambda_{x}k_{x}\epsilon_{p} + i \lambda_{y}k_{y}M) \hat{\sigma}_1 \right. \nonumber\\
&\left. -2 (- \lambda_{y}k_{y}\epsilon_{p} - i \lambda_{x}k_{x}M) \hat{\sigma}_2 - 2ME \hat{\sigma}_3 
 \right] \hat{\sigma}_2,\end{align}
with
\begin{align}
Z=&  -(E^2 - B_{k,M})^2 - 4 
( -\lambda_{x}k_{x}\epsilon_{p} + i \lambda_{y}k_{y}M)^2 \nonumber\\
& - 4(- \lambda_{y}k_{y}\epsilon_{p} - i \lambda_{x}k_{x}M)^{2} - 4 E^2 M^2,
\end{align} and
\begin{align}
B_{k,M}=&\epsilon_{p}^2 + \Delta^2 + \left(\lambda_{x}k_{x}\right)^2 + \left(\lambda_{y}k_{y}\right)^2 - M^2
\end{align}

To discuss the symmetry of the anamolous Green functions, we use the Matsubara representation which can be obtained by analytical continuation, $E + i\delta \to i\omega_n$,
\begin{align}
G_{12} 
=& i \frac{\Delta}{Z} \left[ 
(-\omega_n^2 - B_{k,M}) \hat{\sigma}_0
+2 ( -\lambda_{x}k_{x}\epsilon_{p} + i \lambda_{y}k_{y}M) \hat{\sigma}_1 \right. \nonumber\\
&\left. -2 ( -\lambda_{y}k_{y}\epsilon_{p} - i \lambda_{x}k_{x}M) \hat{\sigma}_2 - i 2M \omega_n \hat{\sigma}_3 
 \right] \hat{\sigma}_2 \\
 \equiv & i( 
f_0 + \boldsymbol{f} \cdot \boldsymbol{\sigma}) \hat{\sigma}_2, \nonumber\\
Z=&  -( - \omega_n^2 - B_{k,M})^2 - 4 
( -\lambda_{x}k_{x}\epsilon_{p} + i \lambda_{y}k_{y}M)^2  \nonumber \\
& - 4(-\lambda_{y}k_{y}\epsilon_{p} - i \lambda_{x}k_{x}M)^{2} + 4 \omega_n^2 M^2,\\
&f_0(\boldsymbol{k},i\omega) = \frac{\Delta}{Z}
(-\omega_n^2 - B_{k,M}),\\
&f_1(\boldsymbol{k},i\omega) = 2 \frac{\Delta}{Z}
 ( -\lambda_{x}k_{x}\epsilon_{p} + i \lambda_{y}k_{y}M),\\
&f_2(\boldsymbol{k},i\omega) = -2 \frac{\Delta}{Z}
  ( -\lambda_{y}k_{y}\epsilon_{p} - i\lambda_{x}k_{x}M),\\
&f_3(\boldsymbol{k},i\omega) = -2i \frac{\Delta}{Z}
 M \omega_n.
\end{align}
The total wave function of a pair of fermions should be asymmetric and the total wave functions  can be described by the product of an orbital (or parity), spin and frequency term. Even-frequency pairing means that a function is even in $\omega_{n}$. If we consider singlet (which is an odd function under spin permutation) $s$-wave (orbitally symmetric) pairing, the pairing wave function $F(=G_{12})$ should satisfy the relation $F_{s}(k)=F_{s}(-k)$ in order for the wave function to be antisymmetric when the pairing is even in frequency. For $p$-wave triplet pairing the relation $F_{p}(k)=-F_{p}(-k)$ applies \cite{Tanaka2007}. 

We note that $Z$ is an even function of $\omega_n$ and is an even-parity function. The singlet component $f_0$ belongs to the even-frequency even-parity class (ESE). The two equal-spin components ($f_1$ and $f_2$ ) are even-frequency spin-triplet odd-parity class (ETO). Finally, $f_3$ belongs to odd-frequency spin-triplet even-parity class (OTE).

For a 3D topological insulator $\epsilon_{p}$ and the SOC strength in $x$ and $y$-direction in Eq. (1) are equal to $\mu$ and the Fermi velocity $v$ respectively. The matrix $[E-H]$ in the relation $[E-H]\textbf{G}=\mathbb{1}$  is then given by
\begin{eqnarray}
\left[\begin{array}{cccc}
E+\mu-M & -v|k|e^{i\theta} & 0 &-\Delta \\
-v|k|e^{-i\theta} & E+\mu+M & \Delta & 0\\
0 & \Delta & E-\mu+M & -v|k|e^{-i\theta}\\
-\Delta & 0 & -v|k|e^{i\theta} & E-\mu-M
\end{array} \right],
\end{eqnarray} where $M$ is a magnetization term and $\theta$ is the angle between $k_{x}$ and $k_{y}$. It is instructive to compare the 3D TI with semiconductors with strong spin-orbit coupling. Although there are analogies between them, also several differences exist between them. For comparison, we show therefore the model for nanowires in the supplementary material.

We assume in this section that $\mu>E_{DP}$ where $DP$ stands for the Dirac point, but the opposite ``hole'' regime of $\mu<E_{DP}$ can also easily be obtained. For a chemical potential far above the Dirac point we can make additionally the assumption that $\mu\gg \Delta$. If we also consider low energy excitations we obtain for the topological insulator
\begin{eqnarray}
G_{12}=\left[\begin{array}{cc}
F_{\upharpoonleft \upharpoonleft}& F_{\upharpoonleft \downharpoonleft}\\
F_{\downharpoonleft \upharpoonleft} & F_{\downharpoonleft \downharpoonleft}
\end{array} \right],
\end{eqnarray} where
\begin{eqnarray}
F_{\upharpoonleft \upharpoonleft}&\sim & 2\Delta |k|ve^{i\theta} \left(M+\mu\right)/Z_{TI}\nonumber,\\
F_{\upharpoonleft \downharpoonleft}&\sim & \Delta \left(M^{2}-\left|k\right|^{2}v^{2}-\mu^{2}\right)/Z_{TI},\nonumber \\
F_{\downharpoonleft \upharpoonleft}&\sim &-\Delta \left(M^{2}-\left|k\right|^{2}v^{2}-\mu^{2}\right)/Z_{TI},\nonumber \\
F_{\downharpoonleft \downharpoonleft}&\sim & 2\Delta |k|ve^{-i\theta} \left(M-\mu\right)/Z_{TI},\\
Z_{TI}&=&-B^{2}_{TI}-4\left(-vk_{x}\mu-ivk_{y}M\right)^2-4\left(-vk_{y}\mu+ivk_{x}M\right)^{2},\nonumber\\
B^{2}_{TI}&=&\left(\mu^{2}+v^{2}|k|^{2}-M^{2}\right)^{2}+2\Delta^{2}\left(\mu^{2}+v^{2}|k|^{2}-M^{2}\right).
\end{eqnarray}$G_{21}$ is related to $G_{12}$ by complex conjugation. The $e^{\pm i\theta}$ factor shows the chiral $p$-wave character of an $s$-wave proximized 3D topological insulator.

For the moment we assume that we have even-frequency symmetry. ($f_{3}$ in Eq. (15) can be neglected in the regime of low energy excitations. The appearance of non-neglible odd-frequency symmetry terms will be discussed in section III.) The second and third relation in Eq. (16), therefore, correspond to $s$-wave pairing and the other two with $p$-wave pairing relations.


The energy dispersion relation of the $s$-wave proximized topological insulator can be obtained by diagonalizing the corresponding Hamiltonian as described above, and is found to be
\begin{eqnarray}
E&=&\pm\sqrt{E_{t'}\pm2\sqrt{E_{s'}}}
\end{eqnarray} where $E_{t'}=M^{2}+\Delta^{2}+\mu^{2}+|k|^{2}v^{2}$ and $E_{s'}=M^{2}\Delta^{2}+M^{2}\mu^{2}+|k|^{2}v^{2}\mu^{2}$. In the limit of $\mu \gg\Delta$ this can in good approximation be written as 
\begin{eqnarray}
E&=&\pm \mu \pm \sqrt{v^{2}|k|^{2}+M^{2}}, \nonumber
\end{eqnarray}
Now we have all the formal work done to consider the pairing symmetry in three different regimes: $|M|=0$, $M<\mu$ and the ``Majorana" regime.

\begin{figure}[t!]
\centering
		\includegraphics[width=0.9\textwidth]{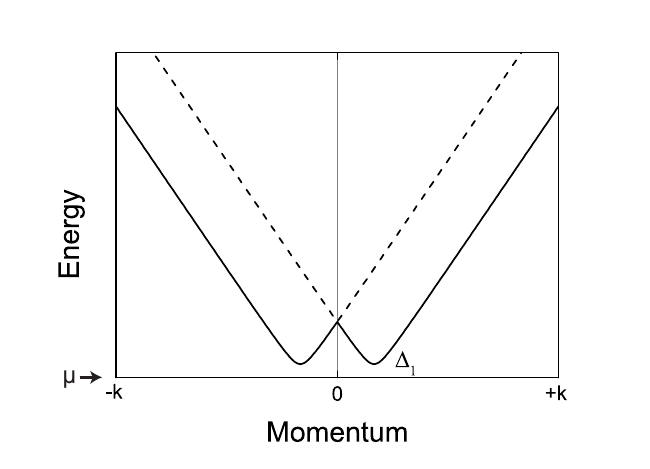}
		\caption{Energy dispersion relation for a 3D topological insulator in the regime $|M|\ll\mu$.  The solid line corresponds to $\sqrt{E_{t'}-2\sqrt{E_{s'}}}$ and the dashed line to $\sqrt{E_{t'}+2\sqrt{E_{s'}}}$. The graph is plotted for $\Delta$ in the order of $\mu$ for clarity.}
		\label{fig:pwave}
		\vspace{-5pt} 
\end{figure}

\subsection{Regime $|M|=0$}
Assume a homogenous magnetization term (for example caused by an external magnetic field). In the regime $|M|\ll\mu$ we have the situation as depicted in Fig. \ref{fig:pwave}. There is a gap opening indicated by $\Delta_{1}$ at positive momentum. At negative momentum a gap with the same size is opening. Using the approached dispersion relation we have $\mu=v|k|$ at the gap opening. Substituting this relation into Eq. (16) we obtain the following anamolous Green function relations at $\Delta_{1}$ 
\begin{eqnarray}
F_{\upharpoonleft \upharpoonleft}\sim 2\Delta |k||k_{f}|e^{i\theta}/Z_{TI} \nonumber,\\
F_{\upharpoonleft \downharpoonleft}\sim -2\Delta |k|^{2}/Z_{TI},\nonumber \\
F_{\downharpoonleft \upharpoonleft}\sim 2\Delta |k|^{2}/Z_{TI},\nonumber \\
F_{\downharpoonleft \downharpoonleft}\sim -2\Delta |k||k_{f}|e^{-i\theta}/Z_{TI} \nonumber.
\end{eqnarray} We see from the above that the matrix elements corresponding to the $s$-wave pairing are in magnitude as large as the matrix elements corresponing to the $p$-wave relations. So in the case of neglible $M$ we have an equal admixture of $s$-wave and $p$-wave correlations. This result is similar to the result found by Tkachov in Ref. \cite{Tkachov2013,Tkachov2013b} for a topological insulator. In Ref. \cite{Fu2008} Fu and Kane transform to another basis, $c_{k}=\left(\psi_{\upharpoonleft k}+e^{i\theta}\psi_{\downharpoonleft k}\right)$, in which the Hamiltonian becomes equivalent to a spinless $p_{x}+ip_{y}$ (dominant $p$-wave). Although this transformation is an elegant way to show that $p$-wave relations are present, when doing predictions for the experimental outcome in this basis, a tranformation also has to be made to the $s$-wave superconductor deposited on top. This would also cause an additional phase shift in the normal superconductor through which one can then conclude that $p$-wave is not longer dominant. Nonetheless, the existing proposals of Fu and Kane for Majorana devices \cite{Fu2008} still hold as it was already noted that breaking time-reversal symmetry can make $p_{x}+ip_{y}$ dominant, as we will see next.  

\subsection{Regime $M < \mu$}
As we turn on the perpendicular magnetization term, it follows from the dispersion relations that we have at $\Delta_{1}$, $\mu^{2}=v^{2}|k|^{2}+M^{2}$ . The anamolous Green function relations become
 \begin{eqnarray}
F_{\upharpoonleft \upharpoonleft}\sim 2\Delta e^{i\theta} \sqrt{\left(\mu-M\right)\left(\mu+M\right)^{3}}/Z_{TI}\nonumber,\\
F_{\upharpoonleft \downharpoonleft}\sim -2\Delta \left(\mu^{2}-M^{2}\right)/Z_{TI},\nonumber \\
F_{\downharpoonleft \upharpoonleft}\sim 2\Delta \left(\mu^{2}-M^{2}\right)/Z_{TI},\nonumber \\
F_{\downharpoonleft \downharpoonleft}\sim -2\Delta e^{-i\theta} \sqrt{\left(\mu-M\right)^{3}\left(\mu+M\right)}/Z_{TI}\nonumber.
\end{eqnarray} As soon as time-reversal symmetry is broken the increase in $M$ causes the $p$-wave component to become larger in magnitude than the $s$-wave component. This can be intuitively understood by notifying that a magnetic field aligns the spins parallel to the field. Therefore spin triplet pairing is favoured above spin singlet pairing. It is therefore possible to already observe dominant $p$-wave features in the regime $M < \mu$.
 
\begin{figure*}[t!]
	\centering 
		\includegraphics[width=0.8\textwidth]{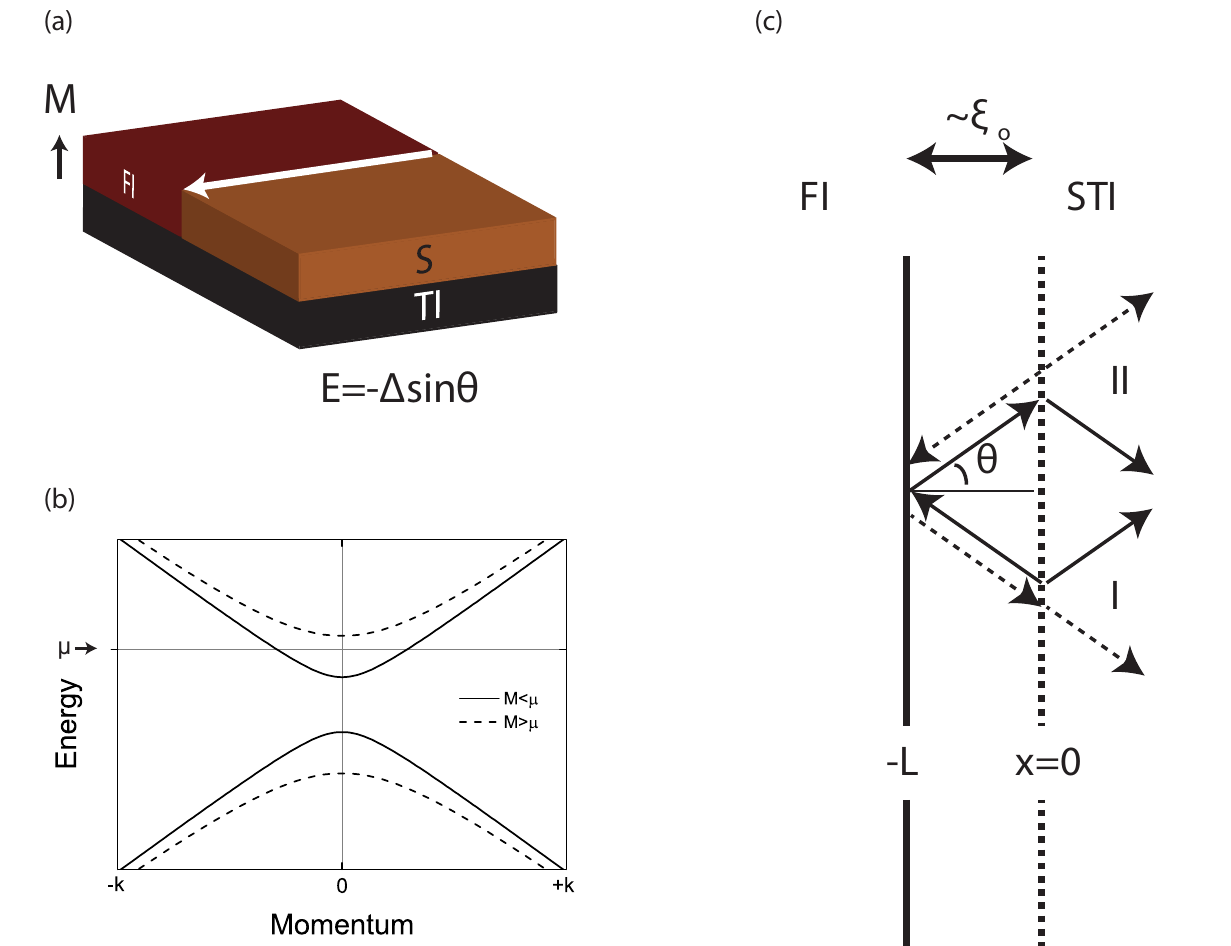}
		\caption{(a) A ferromagnet insulator(FI) can both break time-reversal symmetry and localize the Majorana zero-energy mode in a TI. The location of the Majorana mode is indicated by the white arrow. (b) The dispersion relation of the topological insulator at the ferromagnet side for a magnetization smaller and larger than the chemical potential. In the former, the chemical potential still lays in the Dirac cone. In the latter, the states around the chemical potential are gapped out and the zero-energy mode is localized at the interface between the superconductor and ferromagnet. (c) Calculation of the surface Andreev bound state. To model the suppression of the superconducting gap of the superconducting topological insulator (STI) we assume a normal part in the order of the coherence lenght $\xi_{0}$ at the interface between the superconductor and the ferromagnet.} 
		\label{fig:1}
\end{figure*}
\subsection{The ``Majorana" regime}
Unfortunately, the presence of dominant $p$-wave correlations still does not mean that a Majorana zero-energy mode exists as soon as we apply magnetization, because the zero-energy mode is not yet fully localized. As already note by Fu and Kane \cite{Fu2008}, the 2D TI surface has no edge, which prevents a localized Majorana bound state to form. The same is true in a 2D topological insulator where 1D edge states can go around the 2D topological insulator. We therefore need a different device in order to localize the zero-energy mode.

One way to create an edge \textit{and} break time-reversal symmetry is shown in Fig. \ref{fig:1}(a). The $M$ denotes the magnetization induced by the ferromagnet. We assume here that we have an insulating ferromagnet so that the current is only going through the surface states. A different way is to have a Josephson junction with a phase difference between the two superconductors \cite{Fu2008}. For now, we focus on the device in Fig. \ref{fig:1}(a). In order to have a localized zero-energy mode, the Majorana mode needs to be localized both at the superconducting side and at the ferromagnet side. At the superconducting side the superconducting gap is doing the job. At the ferromagnet side, the magnetization has to be large enough such that the Fermi level is inside the gap (see Fig. \ref{fig:1}(b)). Then, the zero-energy mode is fully localized. To be more specific, in order to have a proper localized Majorana mode one has to satisfy the relation $M(x) > \sqrt{\Delta(x)^{2}+\mu^{2}}$ at the ferromagnet side. We also included the superconducting gap as a function of position to take the proximity effect into account.

\section{Surface Andreev bound states and Majorana zero-energy modes}
So far we have looked at properties of the bulk, i.e. the inner part of the s-wave proximized 2D surface states of the topological insulator. We are now going to discuss the properties at the interface between the superconductor and ferromagnet where the Majorana zero mode is formed in the Majorana regime. It is expected that different symmetries are present at the surface or edge of a superconducting system as noticed in Ref. \cite{Tanaka2012,Gorkov2001}. Due to a finite size, translational symmetry is broken in that direction. The orbital symmetry (even or odd) is then not a well-defined parameter anymore, giving rise to mixed odd and even symmetry at the interface. The spin pairing symmetry, however, is not affected by the finite size. The only option, then, to have an asymmetric wave function is to allow for odd-frequency states at the interface.  

Figure \ref{fig:1}(a) shows the superconductor/ferromagnet device needed to localize a Majorana state. In experiments it is useful to use a ferromagnetic \textit{insulator} (FI) to ensure that the magnet itself does not shunt the device \cite{Linder2010}. We will, therefore, indicate the ferromagnet by FI in the remaining part of the paper. 

Assume, that we are in the regime with a fully localized Majorana state, i.e. $M>\mu$. The resulting SABSs can be calculated by considering a small non-superconducting region in the order of the coherence length to model the suppression of the gap near the edge (Fig. \ref{fig:1}(c)). We use the wave functions described in our previous work (Ref. \onlinecite{Snelder2013}) to calculate the SABSs. The direction of the transmitted waves in the superconductor is chosen such that $k$ is conserved parallel to the interface. We assume that the electrons in the non-superconducting part are confined and therefore, that they completely reflect back at $x=-L$.  

The phase difference that is picked up in one round trip (Fig. \ref{fig:1}(c)) is
\begin{eqnarray}
2L(|k_{e}|-|k_{h}|)-2\theta +3\pi +2\arccos E/\Delta =2\pi n \nonumber \\
=\dfrac{4LE}{\hbar v_{f} \cos \theta}-2\theta + 3\pi + 2\arccos E/\Delta 
\end{eqnarray} so
\begin{eqnarray}
\dfrac{E}{\Delta}=\cos \left(-3\pi/2 +\theta +\dfrac{2LE}{\hbar v_{f} \cos \theta}\right)\\
=-\sin \left(\theta +\dfrac{2LE}{\hbar v_{f} \cos \theta}\right).
\end{eqnarray} Taking the limit $L\rightarrow 0$ we have
\begin{eqnarray}
\dfrac{E}{\Delta}=-\sin (\theta)\\
=-k_{y}/|k|.
\end{eqnarray} When the magnetization of the FI is in the exact opposite direction, the resulting ABS is $\frac{E}{\Delta}=\sin (\theta).$ This is exactly the SABS of a chiral $p$-wave superconductor \cite{Eschrig2010}.\\
\\
In order to determine the features of the SABS and its relation to Majorana zero energy modes, we consider the regimes around $E=0$ and $E \neq 0$ in more detail in the following.

\subsection{Around $E=0$}
In Fig. \ref{fig:1}(c) we can calculate the reflection and transmission coefficients of the electron and holes at the interface by using continuation of the wave function. From these coefficients it follows that the total wave function in the superconducting part in the electron-hole basis $\left[\psi_{k\uparrow}, \psi_{k\downarrow},\psi^{\dagger}_{-k\uparrow}, \psi^{\dagger}_{-k\downarrow}\right]$ is given by
\begin{eqnarray}
\Psi=\dfrac{e^{i\theta}}{\Delta}\left(\begin{array}{c}
(E-i\tilde{\Omega})\\
-e^{-i\theta}(E-i\tilde{\Omega})\\
\Delta e^{-i\theta}\\
\Delta
\end{array}\right)e^{-ikr}e^{-r/\xi_{0}}\nonumber \\
+\dfrac{1}{E+i\tilde{\Omega}}\left(\begin{array}{c}
(E+i\tilde{\Omega})\\
(E+i\tilde{\Omega})e^{i\theta}\\
-\Delta e^{i\theta} \\
\Delta
\end{array}\right)e^{ikr}e^{-r/\xi_{0}},
\end{eqnarray} where $\tilde{\Omega}=\sqrt{\Delta^{2}-E^{2}}$ and $\xi_{0}=\sqrt{\Delta^{2}-E^{2}}/v_{F}$.
Around $E=0$ (and so around $\theta=0$) the expression for the Majorana zero mode becomes
\begin{eqnarray}
\Psi= e^{-x/\xi_{0}}\left(\begin{array}{c}
2e^{-i\pi/4}\cos \left(kx+\pi /4\right)\\
2e^{i\pi/4}\cos \left(kx-\pi / 4\right)\\
2e^{i\pi/4}\cos \left(kx+\pi /4\right)\\
2e^{-i\pi/4}\cos \left(kx-\pi /4\right)\\
\end{array}\right).\label{eq:MFs}
\end{eqnarray} 

Note, that this expression is consistent with the result found by Tanaka and Asano in a semiconductor/superconductor device \cite{Asano2013}. In their case, an additional magnetic field is applied through which only the spin up part survives of Eq. (\ref{eq:MFs}). 
 
To give a complete picture of the properties of these Majorana zero-energy mode, we calculated the frequency symmetry of this zero-energy state. The anamolous Green's function can be obtained by the relation
\begin{eqnarray}
F(E,r,r';\theta)_{\uparrow \downarrow}=\Sigma _{n} \left [ \dfrac{u_{n\uparrow}(r)v_{n\downarrow}^{*}(r')}{E+i\delta -E_{n}}+\dfrac{v_{n\uparrow}^{*}(r)u_{n\downarrow}(r')}{E+i\delta +E_{n}}\right].
\end{eqnarray} where $u$ and $v$ are the electron and hole parts respectively. 

The relationship at $x=x'=0$ satisfies 
\begin{eqnarray}
\overline{F}_{\beta,\alpha}(\textbf{r}, E, \textbf{p})=-\left[F_{\alpha,\beta}(\textbf{r}, -E, \textbf{p})\right]^{*}
\end{eqnarray} where $\overline{F}$ denotes the conjugated of $F$ and $\alpha$ and $\beta$ are the spin-indices\cite{Tanaka2007}, which means the Cooper pairs are odd-frequency pairs. The spatial symmetry of Eq. (\ref{eq:MFs}) satisfies the $s$-wave symmetry. Therefore the boundary of a pure $p$-wave even-frequency superconductor is a pure $s$-wave, odd-frequency state. The appearance of odd-frequency states near zero-energy is also noted in calculations on $p_{x}$ and $p_{x}+ip_{y}$ superconductors. \cite{Bakurskiy2014}

\subsection{$E\neq$ 0}
In a similar way as discussed for zero energy we can calculate the anomalous Green function for energies larger than zero. We find that there is an admixture between even and odd-frequency. The odd and even-frequency parts are, respectively, given by
\begin{eqnarray}
F_{odd}&\backsim &\sqrt{1-E^{2}/\Delta^{2}},\\
F_{even}&\backsim &E/\Delta.
\end{eqnarray} Near zero-energy, the odd-frequency part is dominant and the even-frequency contribution grows with $E$. The emergence of a (partly) odd-frequency amplitude at the surface/interface due to spatial non-uniformity \cite{Tanaka2012,Tanaka2007b,Tanaka2007c,Eschrig2007} or interorbital pairing \cite{BlackSchaffer} is well-known for superconducting systems. However, in this case, we have a full odd-frequency state at zero energy. Therefore the conclusion of Asano and Tanaka \cite{Asano2013} holds that in these devices pure odd-frequency Cooper pairs and the Majorana zero-energy mode are one and the same thing. 
\begin{figure}[t!]
	\centering 
		\includegraphics[width=0.7\textwidth]{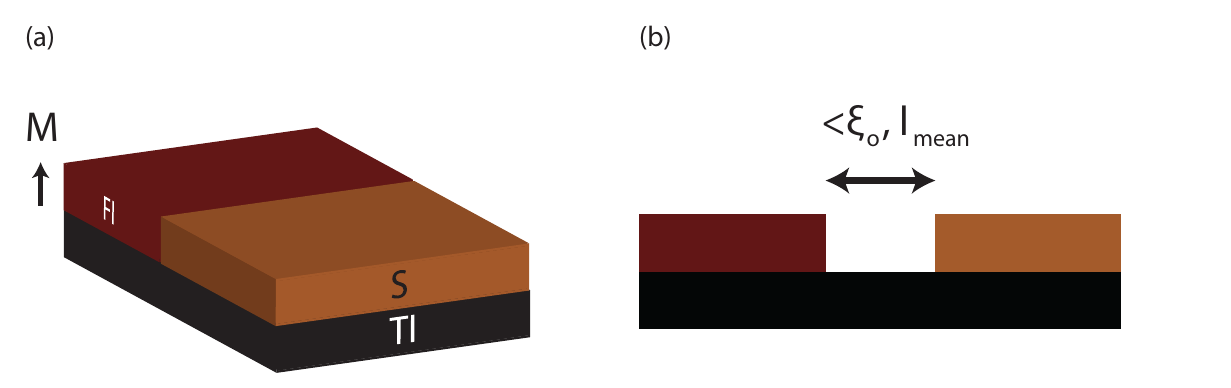}
		\caption{(a) In the calculation of the conductance we consider two electrodes: the proximized TI by the FI and the proximized TI by the $s$-wave superconductor. (b) In order for the Andreev formalism to hold, the distance between the FI and the superconductor should be smaller than the mean free path and the coherence length.} 
		\label{fig:2}
\end{figure} 

\section{TI/STI tunneling conductance}
In the study of the conductance we consider the system where one electrode consists of the topological insulator proximized by the FI and the other electrode of the topological insulator proximized by the $s$-wave superconductor as depicted in Fig. \ref{fig:2}(a). We assume here that the magnetization is smaller than the chemical potential in the magnetic topological insulator electrode. We modelled this magnetic TI/STI interface by using the wave functions in the magnetic TI and STI as described in Ref. \cite{Snelder2013} and match them at the interface such that
\begin{eqnarray}
\psi_{in}+a\psi_{h}+b\psi_{e}&=&t_{e}\psi_{Se}+t_{h}\psi_{Sh}
\end{eqnarray} where $\psi_{h}$ is the reflected hole, $\psi_{e}$ the reflected electron, $\psi_{Se}$ the transmitted quasi-particle in the electron branch of the superconductor, $\psi_{Sh}$ the transmitted quasi-particle in the hole branch of the superconductor and $a$, $b$, $t_{e}$ and $t_{h}$ are the Andreev, normal, electron transmission and hole transmission coefficients respectively. 
We calculated then the conductance by using the relation
\begin{eqnarray}
\dfrac{G}{G_{0}}=\dfrac{\int \limits_{-\pi/2}^{\pi/2}\! d\theta \cos(\theta)T\left(\theta,eV\right)}{\int\limits_{-\pi/2}^{\pi/2}\! d\theta \cos(\theta)T\left(\theta,eV\right)|_{\Delta=0}},
\end{eqnarray} where $T\left(\theta,E\right)=1+|a|^{2}-|b|^{2}$ and $\theta$ is the angle between $k_{x}$ and $k_{y}$. The Andreev formalism holds when the distance between the superconducting and the ferromagnet side is smaller than the coherence length and the mean free path ($l_{mean}$) as drawn in Fig. \ref{fig:2}(b). We choose here to divide the conductance by the energy dependent normal conductance to ensure that the normalized conductance goes to one at large voltage. 

\begin{figure*}[t!]
	\centering 
		\includegraphics[width=0.9\textwidth]{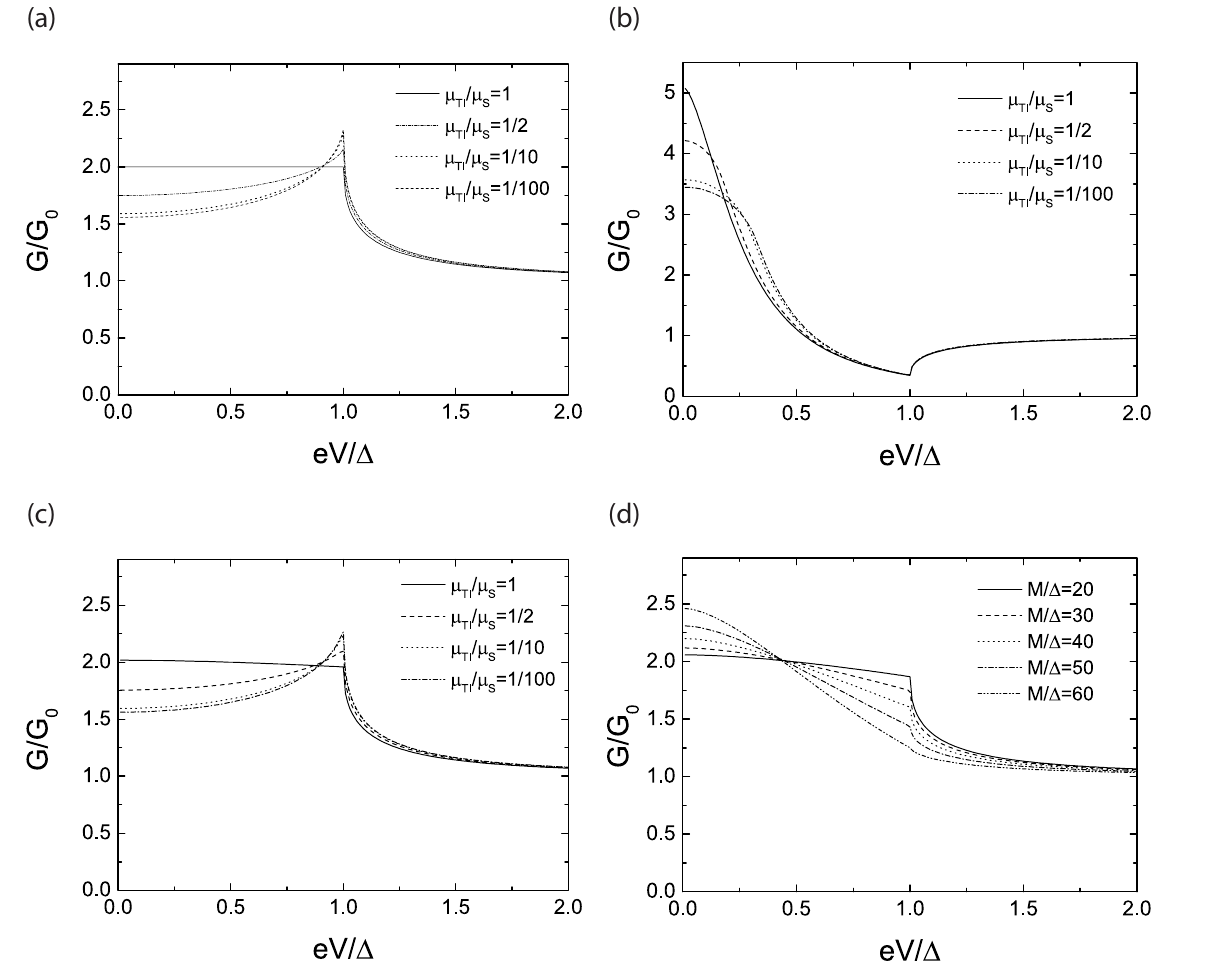}
		\caption{(a) Conductance spectra for a TI/STI configuration for different mismatches in chemical potential. (b) Conductance spectra for a magnetization $M/\Delta$=95, $\mu_{TI}/\Delta$=100 and different mismatches in chemical potential. (c) Conductance spectra for a magnetization $M/\Delta$=10, $\mu_{TI}/\Delta$=100 and different mismatches in chemical potential. (d) The conductance spectra for different values of magnetization and $\mu_{TI}/\Delta=\mu_{S}/\Delta$=100. Even if the magnetization is smaller than the chemical potential, clear signatures of the SABSs are still visible. } 
		\label{fig:3}
\end{figure*}

We first consider the case with time-reversal symmetry, i.e. without the FI on top of the TI. By means of a gate electric field, the chemical potential at the superconductor side can be tuned independently from the non-superconductor side. Therefore, we make from now on a distinction between the chemical potential at the superconductor and non-superconductor side. In Fig. \ref{fig:3}(a) we plotted the normalized conductance for different ratios of the STI chemical potential, $\mu_{S}$, and TI chemical potential, $\mu_{TI}$. A conductance peak at $eV=\Delta$ is appearing for larger barriers due to Andreev resonance at this energy, similar to the s-wave case. Opposite to the s-wave case, the value of the conductance never drops to zero, even for a large mismatch where the barrier height goes practically to infinity ($\mu_{S}/\mu_{T}=1/100$). The conductance inside the gap is close to one instead which can be understood by considering different angles of incidence. The large barrier causes the propagation direction in the superconductor to be almost equal to $\theta=0$ as follows from the conservation of parallel momentum: $\theta_{S}=\arcsin\left(\sin \theta \mu_{TI}/\mu_{S}\right)$. The spin in the topological insulator with $\theta=0$ sees no barrier because of Klein tunneling (which gives perfect Andreev reflection for all energies) and for $\theta=\pi/2$ the spin mismatch is the largest which results in the lowest transparency (perfect normal reflection). For angles inbetween, the transparency of the barrier increases continuously from one to zero for increasing angle of incidence. The angle averaged conductance is therefore close to one. 

The presence of a non-zero conductance for energies below the gap, even for large barriers, distinguishes the proximized TI surface from conventional $s$-wave symmetry superconductors. Although the $p$-wave pairing is not dominant in the time symmetric situation, the $p$-wave correlations are encoded in the non-zero conductance for energies below the gap in the presence of a barrier. The fact that the gap does not go to zero is due to the spin-momentum locking which is also responsible for the $p$-wave correlations. 
\section{TI/STI tunneling conductance with broken time-symmetry}
To modulate the TI/STI tunneling conductance with broken time-reversal symmetry we take the set-up as displayed in Fig. \ref{fig:2}(a). We will use a relatively large chemical potential compared to the superconducting gap to model the experimentally realistic regime. The topological insulators so far have the Dirac point close or at the same energy as the bulk bands so as long as the chemical potential is in the gap, the chemical potential is likely much larger than the superconducting gap. We argued in earlier work\cite{Snelder2013}, that a magnetization in the case of a ferromagnet insulator will only be a few percent of the chemical potential. However, to see the effect of the broken time-symmetry we also modulated the conductance spectrum for a magnetization value of 0.95$\mu_{TI}$. This result is shown in Fig. \ref{fig:3}(b). There are two distinguished features compared to an N/S interface, namely a zero-bias peak (ZBP) and a conductance dip at $eV=\Delta$. The appearance of these two features for large magnetization is also reported by Linder \textit{et al.}\cite{Linder2010}. Both features are due to the formation of SABSs at the interface. In fact, the SABSs have a constant density of states at energies between $-\Delta$ and $+\Delta$ as follows from Eq. (22), but from conservation of parallel momentum it follows that the angles near zero have a larger transparency. The coherence peaks at the gap energy in the low barrier limit, are replaced by resistance peaks at the same energy. The coherence dips at $\Delta$ arise from the spectral weight transfer from high to low energy due to the formation of SABSs. This results in a decrease of the density of states at $\Delta$. Therefore, next to a zero bias conductance peak, also coherence dips are a distinguished feature of a $p$-wave order parameter compared to $s$-wave. 

Note, that the surface states are not fully gapped yet with a magnetization of 95$\%$ of the chemical potential, i.e. the chemical potential is not inside the gap at the FI side. We are therefore not in the topologically non-trivial regime. However, the increased reflecion at the interface or enhanced $p$-wave pairing symmetry due to magnetization, causes the SABS to form at the interface. If we use a magnetization value of 10$\%$ of the chemical potential in the TI, Fig. \ref{fig:3}(c) is obtained. We see there is only a small effect on the conductance spectra around zero energy at $\Delta$ compared to the time-reversal symmetry case in Fig. \ref{fig:3}(a).  Figure \ref{fig:3}(d) shows the effect of increasing the magnetization for $\mu_{TI}=\mu_{S}=100\Delta$. Gradually, the conductance at zero energy increases while at $\Delta$ the conductance decreases due to the formation of SABSs for larger magnetization. It shows that the ZBP and conductance dip at the gap energy come together. That is, if one has a dominant $p$-wave pairing wave function and if one observes a ZBP, then also the conductance dips are present. As soon as the ZBP becomes less pronounced, as is the case for lower magnetizations, the conductance dip at $eV=\Delta$ is less pronounced. The observation of \textit{both} a ZBP and a conductance dip is a strong signature for a $p$-wave state. For practical ferromagnets that have a magnetization in the order of a few percent of the chemical potential of the TI, it is not expected to see this particular $p$-wave signatures back in the conductance spectra. A pronounced deviation from $s$-wave behaviour starts to occur when $M/\Delta=60$ (see Fig. \ref{fig:3}(d)). With a typical exchange energy of 40 meV \cite{Semenov2012,Kong2011} it means that one has to gate tune the chemical potential to about 70 meV from the Dirac point, which seems doable experimentally.\cite{Kim2012,Yang2014}  


\section{Discussion and conclusion}
An equal admixture of $s$ and $p$-wave correlations exists in a 3D topological insulator proximized by an $s$-wave superconductor. By inducing a perpendicular magnetization, the $p$-wave pairing becomes dominant. For a magnetization energy as large as the chemical potential, the topological regime is entered with Majorana modes.

We studied the symmetry of the bound state of a 3D topological insulator in the Majorana regime. Green function techniques show us that the pure odd-frequency pairing state at interfaces is equivalent to a zero-energy Majorana mode.  

In the conductance spectra the increasing $p$-wave pairing can be observed by the presence of a ZBP together with a conductance dip at the gap energy. In the time-reversal symmetric case, the non-zero sub-gap conductance in the presence of large interface barriers, indicates that $p$-wave correlations are present. \\
\\
We acknowledge insightful discussions with M. P. Stehno, Y. Tanaka and E. M. Hankiewicz. This work is supported by the Netherlands Organization for Scientific Research (NWO), by the Dutch Foundation for Fundamental Research on Matter (FOM), the European Research Council (ERC) and supported in part by Ministry of Education and Science of the Russian Federation, grant no. 14Y26.31.0007. 

\end{document}


\title{Supplementary material for  ``Observability of surface Andreev bound states in a topological insulator in proximity to an s-wave superconductor"}
\author{M. Snelder}
\affiliation{Faculty of Science and Technology and MESA+ Institute for Nanotechnology, University of Twente, 7500 AE Enschede, The Netherlands}
\author{A. A. Golubov}
\affiliation{Faculty of Science and Technology and MESA+ Institute for Nanotechnology, University of Twente, 7500 AE Enschede, The Netherlands}
\affiliation{Moscow Institute of Physics and Technology, Dolgoprudny, Moscow 141700, Russia}
\author{Y. Asano}
\affiliation{Department of Applied Physics and Center for Topological Science \& Technology, Hokkaido University, Sapporo 060-8628, Japan}
\author{A. Brinkman}
\affiliation{Faculty of Science and Technology and MESA+ Institute for Nanotechnology, University of Twente, 7500 AE Enschede, The Netherlands}

\date{\today}
\maketitle
In this Supplementary we review the superconducting symmetry in nanowires with strong spin-orbit coupling as a comparison to the discussion on the 3D TI in the main text. A single spin branch is now only achieved after applying a Zeeman magnetic field. We review in which regime a dominant $p$-wave correlation is present and under which conditions a dominant $p$-wave correlation can lead to a zero-energy Majorana bound state. 

\section{Pairing wave function and Majorana-modes in nanowires}
The chemical potential is usually close to the conduction band so that $\mu\gg \Delta$. If we also consider low energy excitations we obtain for the nanowire ($k_{x}=0, \lambda_{y}=\lambda$)
\begin{eqnarray}
F_{\upharpoonleft \upharpoonleft}&\sim & -2\Delta i k \lambda \left(M+\dfrac{\hbar^{2} k^{2}}{2m}-\mu\right)/Z_{nw}\nonumber,\\
F_{\upharpoonleft \downharpoonleft}&\sim & \Delta \left(M^{2}-k^{2}\lambda^{2}-\left(\dfrac{\hbar^{2}k^{2}}{2m}\right)^{2}+\dfrac{\hbar^{2}k^{2}}{m}\mu-\mu^{2}\right)/Z_{nw},\nonumber \\
F_{\downharpoonleft \upharpoonleft}&\sim & -\Delta \left(M^{2}-k^{2}\lambda^{2}-\left(\dfrac{\hbar^{2}k^{2}}{2m}\right)^{2}+\dfrac{\hbar^{2}k^{2}}{m}\mu-\mu^{2}\right)/Z_{nw},\nonumber \\
F_{\downharpoonleft \downharpoonleft}&\sim & 2\Delta i k \lambda \left(M-\dfrac{\hbar^{2} k^{2}}{2m}+\mu\right)/Z_{nw},\label{Pairing}\\
Z_{nw}&=&-B^{2}_{nw}-4\left(i\lambda k M\right)^{2}-4\left(\lambda k \epsilon_{p}\right)^{2}\nonumber\\
B^{2}_{nw}&=&\left(\epsilon_{p}^{2}+\lambda^{2}k^{2}-M^{2}\right)^{2}+2\Delta^{2}\left(\epsilon_{p}^{2}+\lambda^{2}k^{2}-M^{2}\right). 
\end{eqnarray}  

The energy dispersion relation of the $s$-wave proximized nanowire can be obtained by diagonalizing the corresponding Hamiltonian as described above, and is found to be
\begin{eqnarray}
E &=& \pm\sqrt{E_{t}\pm 2\sqrt{E_{s}}},
\end{eqnarray}
where $E_{t}=\Delta^{2}+\epsilon_{p}^{2}+\lambda^{2}k^{2}+M^{2}$ and $E_{s}=\epsilon_{p}^{2}\lambda^{2}k^{2}+\Delta^{2}M^{2}+\epsilon_{p}^{2}M^{2}$. In the limit of $\mu \gg\Delta$ this can in good approximation be written as 
\begin{eqnarray}
E=\pm \epsilon_{p}\pm\sqrt{\lambda^{2}k^{2}+M^{2}}, \nonumber
\end{eqnarray}
to calculate the pairing symmetry in three different regimes $M \ll \sqrt{\Delta^{2}+\mu^{2}}$, $M =\sqrt{\Delta^{2}+\mu^{2}}$ and $M \gg \sqrt{\Delta^{2}+\mu^{2}}$.

\begin{figure}[t!]
\centering
		\includegraphics[width=0.9\textwidth]{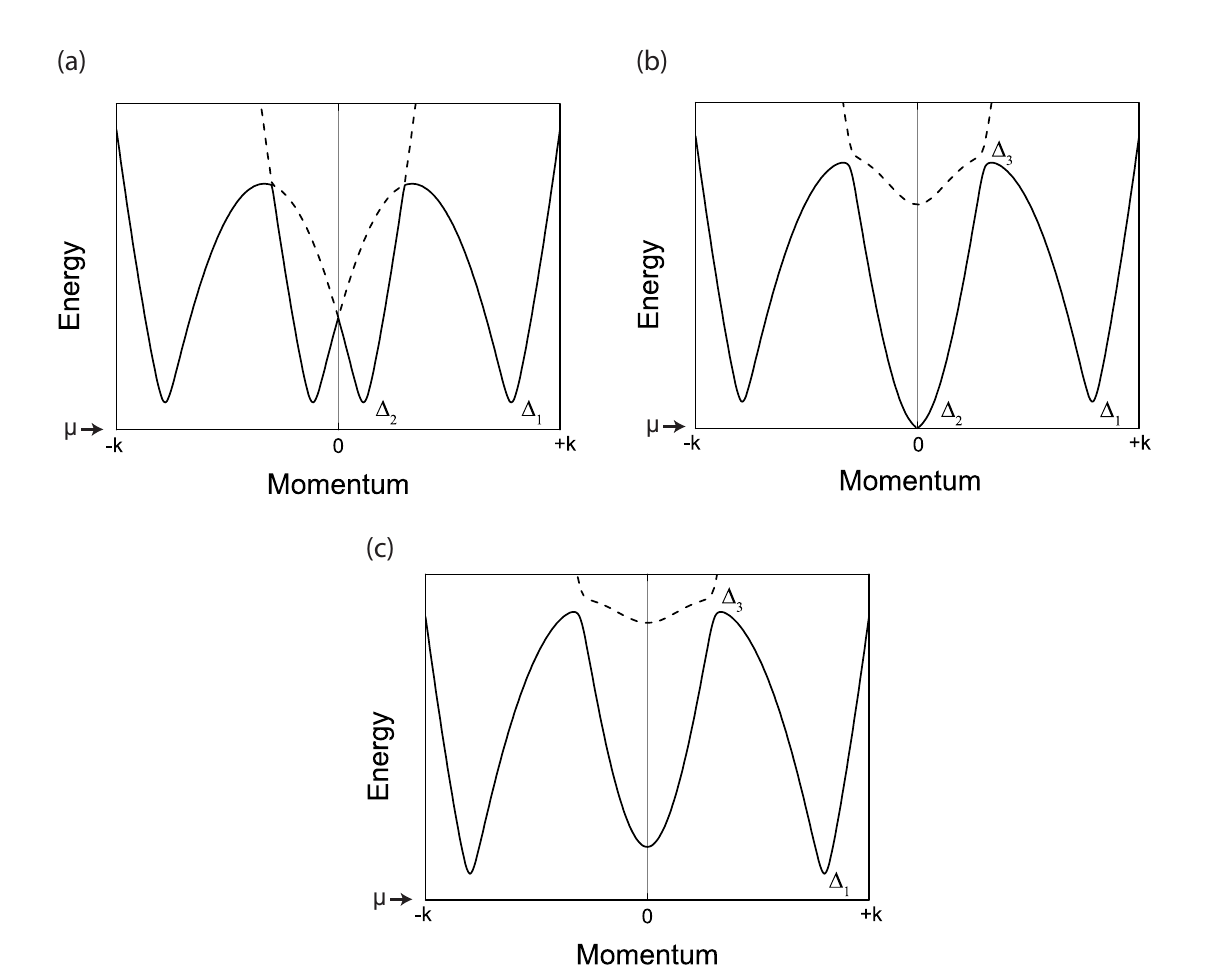}
		\caption{(a) Energy dispersion relation for a nanowire with $M \ll \sqrt{\Delta^{2}+\mu^{2}}$. The solid line corresponds to the relation $\sqrt{E_{t}-2\sqrt{E_{s}}}$ and the dashed line to $\sqrt{E_{t}+2\sqrt{E_{s}}}$. (b) $M=\sqrt{\Delta^{2}+\mu^{2}}$ the gap at $k=0$ is closing. Because of the finite magnetic field there is another gap indicated by $\Delta_{3}$. (c) In the regime $M \gg \sqrt{\Delta^{2}+\mu^{2}}$ a dominant topological non-trivial $p$-wave gap exist at $\Delta_{1}$. The graphs are plotted for $\Delta$ in the order of $\mu$ for clarity.}
		\label{fig:pwave}
		\vspace{-5pt} 
\end{figure}

\subsection{Regime $M \ll \sqrt{\Delta^{2}+\mu^{2}}$}
In this case we have four superconducting gaps as indicated in Fig. \ref{fig:pwave}(a). $\Delta_{1}$ and $\Delta_{2}$ indicate the position for positive wave vector momentum. At $\Delta_{1}$ (and for the corresponding gap at negative $k$) it follows from the dispersion relation that  $\mu=\frac{\hbar^{2}k_{F}^{2}}{2m}-\lambda |k_{F}|$ and for $\Delta_{2}$ $\mu=\frac{\hbar^{2}k_{F}^{2}}{2m}+\lambda |k_{F}|$. Using the relations for the anomolous Green function, Eq. (\ref{Pairing}), we get for $\Delta_{1}$
\begin{eqnarray}
F_{\upharpoonleft \upharpoonleft}\sim -2\Delta i kk_{f} \lambda^{2}/Z_{nw} \nonumber,\\
F_{\upharpoonleft \downharpoonleft}\sim -2\Delta\lambda^{2}k^{2}/Z_{nw},\nonumber \\
F_{\downharpoonleft \upharpoonleft}\sim 2\Delta\lambda^{2}k^{2}/Z_{nw},\nonumber \\
F_{\downharpoonleft \downharpoonleft}\sim 2\Delta i kk_{f} \lambda^{2}/Z_{nw}\nonumber.
\end{eqnarray} Expressions such as $\frac{\hbar^{2}k^{2}}{2m}-\frac{\hbar^{2}k_{F}^{2}}{2m}$ are set to zero in the above equations as we consider low energy excitations. For $\Delta_{2}$ we obtain
\begin{eqnarray}
F_{\upharpoonleft \upharpoonleft}\sim 2\Delta i kk_{f} \lambda^{2}/Z_{nw} \nonumber,\\
F_{\upharpoonleft \downharpoonleft}\sim -2\Delta\lambda^{2}k^{2}/Z_{nw},\nonumber \\
F_{\downharpoonleft \upharpoonleft}\sim 2\Delta\lambda^{2}k^{2}/Z_{nw},\nonumber \\
F_{\downharpoonleft \downharpoonleft}\sim -2\Delta i kk_{f} \lambda^{2}/Z_{nw}\nonumber.
\end{eqnarray} 
We see also here that in the case of time-reversal symmetry there is an equal admixture of $s$ and $p$-wave correlations.  



\subsection{Regime $M=\sqrt{\Delta^{2}+\mu^{2}}$}
This case is shown in Fig. \ref{fig:pwave}(b), where the superconducting gap at $k=0$ is closed. At $\Delta_{2}$ the anomolous Green function parts are zero because $k=0$. Because we have turned on a magnetic field, the bands corresponding to different helicity are not orthogonal anymore. Therefore, there exist an interaction between the two in this regime and another superconducting gap is opening above $\mu$ ($\Delta_{3}$ in the Fig. \ref{fig:pwave}(b)).\cite{SanJose2013} For $\Delta_{3}$  the relation $\mu=\frac{\hbar^{2}k^{2}}{2m}$ holds so that 
\begin{eqnarray}
F_{\upharpoonleft \upharpoonleft}\sim -2\Delta i k\lambda\mu/Z_{nw} \nonumber,\\
F_{\upharpoonleft \downharpoonleft}\sim -\Delta\left(k^{2}\lambda^{2}-\mu^{2}\right)/Z_{nw},\nonumber \\
F_{\downharpoonleft \upharpoonleft}\sim \Delta\left(k^{2}\lambda^{2}-\mu^{2}\right)/Z_{nw},\nonumber \\
F_{\downharpoonleft \downharpoonleft}\sim 2\Delta i k\lambda \mu/Z_{nw}\nonumber.
\end{eqnarray}
For a typical \cite{Mourik2012} spin orbit strength of $\lambda=0.2$ eV$\cdot$\AA\ and $m=0.015m_{e}$ where $m_{e}$ is the free electron mass, it follows that the magnitude of the spin-singlet order parameters in this regime are larger than the $p$-wave components. However, since the energy level of $\Delta_{3}$ increases with $M$, i.e. $E_{\Delta_{3}}\approx \sqrt{\lambda^{2}k^{2}+M^2}$, for low energy excitations this superconducting contribution can be neglected. 

At $\Delta_{1}$, $\frac{\hbar^{2}k_{F}^{2}}{2m}-\mu-\sqrt{\lambda^{2}k_{F}^{2}+\mu^{2}}=0$ so that we get the following relations for the anamolous Green function part:
\begin{eqnarray}
F_{\upharpoonleft \upharpoonleft}\sim -2\Delta i k\lambda\left(\mu+\sqrt{\lambda^{2}k_{f}^{2}+\mu^{2}}\right)/Z_{nw}\nonumber,\\
F_{\upharpoonleft \downharpoonleft}\sim -2\Delta\lambda^{2}k^{2}/Z_{nw},\nonumber \\
F_{\downharpoonleft \upharpoonleft}\sim 2\Delta\lambda^{2}k^{2}/Z_{nw},\nonumber \\
F_{\downharpoonleft \downharpoonleft}\sim 2\Delta i k\lambda\left(\mu-\sqrt{\lambda^{2}k_{f}^{2}-\mu^{2}}\right)/Z_{nw}\nonumber.
\end{eqnarray} Here, we see that $F_{\downharpoonleft \downharpoonleft}$ becomes gradually smaller and $F_{\upharpoonleft \upharpoonleft}$ becomes larger than the $s$-wave pairing parts, assuming $M$ to be positive. We obtain a dominant $p$-wave component for the spin triplet component where the spins are parallel to the magnetic field. 

\subsection{Regime $M \gg \sqrt{\Delta^{2}+\mu^{2}}$, ``Majorana" regime}
The situation is sketched in Fig. \ref{fig:pwave}(c). The gap at $k=0$ reopens. The pairing wave function at $\Delta_{1}$ satisfies
\begin{eqnarray}
F_{\upharpoonleft \upharpoonleft}\sim -2\Delta i k\lambda\left(M+\sqrt{\lambda^{2}k_{f}^{2}+M^{2}}\right)/Z_{nw} \nonumber,\\
F_{\upharpoonleft \downharpoonleft}\sim -2\Delta\lambda^{2}k^{2}/Z_{nw},\nonumber \\
F_{\downharpoonleft \upharpoonleft}\sim 2\Delta\lambda^{2}k^{2}/Z_{nw},\nonumber \\
F_{\downharpoonleft \downharpoonleft}\sim 2\Delta i k\lambda\left(M-\sqrt{\lambda^{2}k_{f}^{2}+M^{2}}\right)/Z_{nw}\nonumber.
\end{eqnarray} We see that for increasing $M$ the $p$-wave part becomes more pronounced and that $F_{\upharpoonleft \upharpoonleft}$ becomes larger than $F_{\downharpoonleft \downharpoonleft}$. $F_{\downharpoonleft \downharpoonleft}$ actually goes to zero for increasing positive field. For negative field $F_{\downharpoonleft \downharpoonleft}$ will be dominant as one would expect for the opposite field direction. Around $k=0$ there is no interaction anymore between the hole and electron branches and, as noted before, the degeneracy in the $p$-wave order parameters disappears in this regime. 

At $\Delta_{3}$ we have
\begin{eqnarray}
F_{\upharpoonleft \upharpoonleft}\sim -2\Delta i k\lambda M/Z_{nw} \nonumber,\\
F_{\upharpoonleft \downharpoonleft}\sim 2\Delta\left(M^{2}-k^{2}\lambda^{2}\right)/Z_{nw},\nonumber \\
F_{\downharpoonleft \upharpoonleft}\sim -2\Delta\left(M^{2}-k^{2}\lambda^{2}\right)/Z_{nw},\nonumber \\
F_{\downharpoonleft \downharpoonleft}\sim 2\Delta i k\lambda M/Z_{nw}\nonumber.
\end{eqnarray}
Hence, the $s$-wave pairing becomes dominant for increasing magnetic field at $\Delta_{3}$, but as discussed before it is safe to ignore it in the low excitation regime.\\
\\
Since the gap at $k=0$ closed for $M=\sqrt{\Delta^{2}+\mu^{2}}$ and has now reopened again we also have a proper localized single Majorana mode at the ends of the nanowire. Just as the surface states of a topological insulator are localized to a surface, the Majorana mode in a nanowire is localized at the ends of the nanowire due to the vacuum gap at one side and the superconducting gap with a reversed band order on the other side. The reversed band order ensures that the bands are crossing at the ends of the nanowire resulting in a localized end mode similar to the topological surface and edge states in topological insulators. \\
\\
Note, that all of the expressions in this section can be generalized to two dimensions. This could be relevant for InAs thin films for example. The scalar $k$ has to be replaced by a $\textbf{k}$-vector with an amplitude $|\textbf{k}|$ and phase $e^{i\theta}$ analogous to the 3D topological insulator discussed below. The $e^{i\theta}$ factor denotes the chirality of one branch and  $e^{-i\theta}$ the chirality of the other branch present at the Fermi level. Apart from an additional $e^{i\theta}$ and $e^{-i\theta}$ factor in the expressions of the order parameter, the given conclusions would still be valid.
